\documentclass[a4paper,twocolumn,DIV=14]{scrartcl}

\pdfoutput=1

\usepackage[utf8]{inputenc}
\usepackage{fourier}
\usepackage[T1]{fontenc}

\usepackage{scrlayer-scrpage}
\lohead{M. Pflüger et al.: Distortion analysis of crystalline and locally quasicrystalline 2D photonic structures with GISAXS}
\usepackage{sidecap}
\sidecaptionvpos{figure}{t}
\usepackage{afterpage}
\usepackage{graphicx}
\usepackage{amsmath}
\usepackage{amssymb}
\usepackage[exponent-product=\cdot,scientific-notation=false,separate-uncertainty=false,per-mode=symbol]{siunitx}
\usepackage{textcomp}

\usepackage{todonotes}

\usepackage{natbib}
\setcitestyle{authoryear}
\newcommand{\citeasnoun}{\citet}
\renewcommand{\cite}{\citep}

\usepackage{bm}

\newcommand{\ZZ}{\mathbb{Z}}
\newcommand{\FF}{\mathcal{F}}
\newcommand{\CS}{\mathcal{S}}

\DeclareMathOperator{\round}{round}

\usepackage{doi}   

\usepackage[format=plain,indention=5pt,labelfont=bf,figurename=Figure,tablename=Table]{caption} 

\usepackage[protrusion=true,expansion=true]{microtype} 

\usepackage{hyperref}

\hypersetup{
    pdfauthor={Mika Pflüger, Victor Soltwisch, Jolly Xavier, Jürgen Probst, Frank Scholze, Christiane Becker, Michael Krumrey},     
    pdfkeywords={GISAXS, Grazing Incidence Small Angle X-Ray Scattering, quasicrystals, lattice misalignment, photonic crystals},
    pdftitle={Distortion analysis of crystalline and locally quasicrystalline 2D photonic structures with GISAXS},
    pdfsubject={We use GISAXS to examine crystalline and locally quasicrystalline 2D photonic structures produced by nanoimprint lithography for photovoltaic applications.
Using a hierarchical theoretical description, we quantify lattice distortions from the GISAXS measurements, showing the differences in reproduction quality between the investigated samples.
},
    colorlinks=true,       
    linkcolor=black,          
    citecolor=black,          
    filecolor=black,          
    urlcolor=blue,            
    menucolor=black,          
}

\begin{document}

\title{Distortion analysis of crystalline and locally quasicrystalline 2D photonic structures with GISAXS}

\author{Mika Pflüger\thanks{Contact: mika.pflueger@ptb.de} \thanks{Physikalisch-Technische Bundesanstalt (PTB), Abbestraße 2-12, 10587 Berlin, Germany} \and Victor Soltwisch\footnotemark[2] \and Jolly Xavier  \thanks{Helmholtz-Zentrum Berlin (HZB), Albert-Einstein-Straße 15, 12489 Berlin, Germany} \and Jürgen Probst\footnotemark[3] \and Frank Scholze\footnotemark[2] \and Christiane Becker\footnotemark[3] \and Michael Krumrey\footnotemark[2]}

\maketitle


\begin{abstract}
In this study, grazing incidence small-angle X-ray scattering (GISAXS) is used to collect statistical information on dimensional parameters in an area of \SI{20x15}{mm} on photonic structures produced by nanoimprint lithography.
The photonic structures are composed of crystalline and locally quasicrystalline two-dimensional patterns with structure sizes between about \SI{100}{nm} and \SI{10}{\micro\metre} to enable broadband visible light absorption for use in solar energy harvesting.
These first GISAXS measurements on locally quasicrystalline samples demonstrate that GISAXS is capable of showing the locally quasicrystalline nature of the samples while at the same time revealing the long-range periodicity introduced due to the lattice design.
We describe the scattering qualitatively in the framework of the distorted wave Born approximation using a hierarchical model mirroring the sample design, which consists of a rectangular and locally quasicrystalline supercell which is repeated periodically to fill the whole surface.

The nanoimprinted samples are compared to a sample manufactured using electron beam lithography and the distortions of the periodic and locally quasiperiodic samples are quantified statistically.
Due to the high sensitivity of GISAXS to deviations from the perfect lattice, the misalignment of the crystallographic axes was measured with a resolution of \ang{0.015}, showing distortions up to $\pm \ang{0.15}$ in the investigated samples.

\end{abstract}

\section{Introduction}
\label{sec:intro}

Grazing incidence small-angle X-ray scattering (GISAXS) is a powerful non-destructive technique for the investigation of nanostructured surfaces \cite{renaud_2009_ProbingSurface,hexemer_2015_AdvancedGrazingincidence}.
It has been used, for example, to measure the morphology of active layers in organic photovoltaics \cite{gu_2012_MultiLengthScaleMorphologies,muller-buschbaum_2014_ActiveLayer,muller-buschbaum_2016_GISAXSGISANS}, the defects in photonic crystals \cite{zhou_2012_LargeareaCrackfree}, interface roughness of layered systems \cite{holy_1993_XrayReflection,babonneau_2009_WaveguidingCorrelated} and lithographically produced nanostructures \cite{jones_2003_SmallAngle,soccio_2014_AssessmentGrazingincidence,soltwisch_2017_ReconstructingDetailed,hagihara_2017_CapabilityMeasuring}.
Local measurement techniques like scanning probe microscopy (e.g. scanning electron microscopy and atomic force microscopy) offer nanometre resolution and detailed insights into the investigated nanostructures, but only measure comparatively small areas (\si{\micro\meter\squared}).
Collecting statistical information on large nanostructured surfaces is therefore prohibitively time-consuming.
In contrast, visible light microscopy, scatterometry and spectroscopy can quickly measure large areas, but do not easily offer nanometre resolution.
Bridging this gap, GISAXS provides statistical information on the dimensional properties over the whole measurement area of about \SI{10}{\milli\meter\squared} with nanometre resolution.
This enables the investigation of imperfections in otherwise highly ordered lithographically produced nanostructures \cite{rueda_2012_GrazingincidenceSmallangle,soccio_2014_AssessmentGrazingincidence,soltwisch_2016_CorrelatedDiffuse,suh_2016_CharacterizationShape} even for buried structures \cite{gann_2014_TopographicMeasurement} as well as the investigation of the increasing disorder in nanostructures during annealing \cite{meier_2012_SituFilm}.
However, GISAXS as a scattering technique does not offer direct imaging which necessitates rather complex data analysis and interpretation.

In this study, we use nanoimprint lithography, and for comparison electron beam lithography, to manufacture periodic and locally quasiperiodic two-dimensional photonic structures intended for use in the solar spectrum, and investigate the samples using GISAXS.
To minimize the resource and energy usage of silicon photovoltaics, it is desirable to use thin active silicon films.
In order to maintain high efficiency in thin-film devices, the active film is structured over the whole device surface with a structure size ranging from about \SI{100}{nm} to \SI{10}{\micro\metre}, enhancing the absorbance using light trapping or wave-optics effects \cite{becker_2013_PolycrystallineSilicon,priolo_2014_SiliconNanostructures}.
In the finished device, the structured interface is buried under other layers (the active film, front contact or coating layers) \cite{eisenhauer_2015_GrazingIncidence}, which makes it difficult to measure.
Here, GISAXS provides an attractive measurement technique due to the possibility to vary the depth sensitivity using different incident angles and incident photon energies \cite{jiang_2011_WaveguideenhancedGrazingincidence,wernecke_2014_DepthDependentStructural}.

Local quasicrystallinity provides a structuring approach which leads to broadband high absorption \cite{xavier_2014_QuasicrystallinestructuredLight,xavier_2016_DeterministicComposite}.
Quasicrystals are ordered, non-periodic structures with an essentially sharp diffraction pattern, which often exhibits non-crystallographic rotational symmetry \cite{shechtman_1984_MetallicPhase,iucr_2018_Quasicrystal}.
The Fourier spectrum of quasiperiodic lattices is dense, but the Fourier components are sharp and differ in intensity, so that the scattering image shows sharp diffraction orders \cite{suck_2002_QuasicrystalsIntroduction}.
We use the term \emph{locally quasicrystalline} to describe structures satisfying the following conditions. Firstly, they must exhibit an essentially sharp diffraction pattern with a non-crystallographic rotational symmetry. Secondly, the structures are nevertheless periodic with a large unit cell.
Locally quasicrystalline structures combine the optical properties of quasicrystals with periodicity, which is beneficial for manufacturing \cite{xavier_2014_QuasicrystallinestructuredLight}.

As a first step towards the characterization of complex buried interfaces found in modern solar-cell devices, we report on GISAXS measurements and simulations of locally quasiperiodic surface structures.
We find that, due to the very different resolutions in the scattering plane and perpendicular to it, both the local quasicrystallinity and the long-range periodicity are visible in the GISAXS patterns.
The long-range periodicity restricts scattering to semicircles at fixed $q_x$, but the local quasiperiodicity leads to non-periodic spacing of the diffraction orders and a correspondingly rich power spectral density on the scattering semicircles.
Furthermore, we compare periodic and locally quasiperiodic samples produced using nanoimprint lithography with a reference sample produced using electron beam lithography.
We find that GISAXS is very sensitive to local distortions of the periodicity of the tiling, revealing that deviations up to $\pm\ang{0.15}$ were introduced in the nanoimprinted samples.
These distortions are sufficiently small to not impair the functioning of the nanophotonic device at the intended wavelengths larger than \SI{250}{nm}.

\section{Methods}

\subsection{Sample preparation}
\label{ssec:samples}

We investigated nanostructured silicon oxide samples consisting of \SI{275}{\nano\meter} high pillars forming a crystalline or locally quasicrystalline structure produced by nanoimprint lithography (NIL) \cite{verschuuren_2007_3DPhotonic}, and for reference, a comparable silicon sample with \SI{370}{\nano\meter} high pillars forming a crystalline structure manufactured using electron-beam lithography (EBL) (see table \ref{tab:samples}).
Quasicrystalline lattices can be obtained by an approach based on Fourier reconstruction published in \citeasnoun{xavier_2010_ReconfigurableOptically}.
However, for the production of the samples by EBL and NIL, the lattice design must be periodic.
To retain the optical properties of the quasicrystalline structures, we therefore use lattices comprised of tiled rectangular supercells with local quasicrystallinity as described in \citeasnoun{xavier_2014_QuasicrystallinestructuredLight}.
In the 10-fold rotationally symmetric locally quasicrystalline structure of the \emph{NIL 10-fold quasi} sample, the supercell covers a \SI{9.4x11.05}{\micro\meter} area and contains 271 discrete lattice points, and in the \emph{NIL 12-fold quasi} sample, the supercell covers \SI{10.8x13.6}{\micro\metre} and contains 612 discrete lattice points.

\begin{table*}[t]
\small
\caption{Samples investigated in this study\label{tab:samples}}
\begin{tabular}{lllccc}
Sample                   & Lithography & Periodicity   & Rotational symmetry     & Nearest neighbour & Pillar\\
                         &             &               & of diffraction pattern  & distance / nm     & diameter / nm\\
\hline
\emph{EBL}               & E-beam      & periodic      & 4-fold                  & 700               & 300 \\
\emph{NIL hexagonal}     & nanoimprint & periodic      & 6-fold                  & 802               & 293 \\
\emph{NIL 10-fold quasi} & nanoimprint & locally       & 10-fold                 & 510               & 257 \\
                         &             & quasiperiodic &                         &                   & \\
\emph{NIL 12-fold quasi} & nanoimprint & locally       & 12-fold                 & 405               & 216 \\
                         &             & quasiperiodic &                         &                   & \\
\end{tabular}
\end{table*}

In EBL, a photo resist is coated onto a silicon substrate and exposed with a pattern using an electron beam.
The developed pattern is transferred to a deposited nickel mask via lift-off and finally into the substrate via etching.
For NIL, a master structure is fabricated with EBL and a transfer negative of the master structure is taken.
The transfer negative is used as a mould for pattern replication by pressing it into a sol-gel coated onto a substrate.
Subsequent hardening of the sol-gel results in the investigated silicon oxide samples.
Details of all the used ingredients and procedures for NIL and EBL are described in \citeasnoun{xavier_2016_DeterministicComposite}.
The NIL production process was not particularly optimized for the highest reproduction quality, since one aspect of the present study is the quantification of inhomogeneities.

\subsection{GISAXS}
\begin{SCfigure*}[1.0][tb]
\includegraphics[width=1.2\columnwidth]{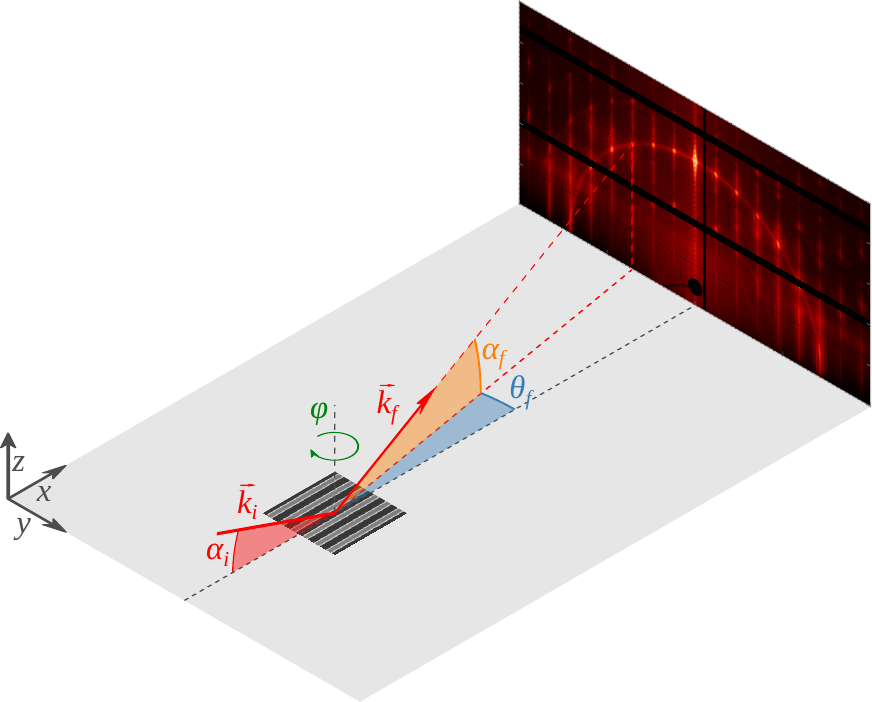}
\caption{Geometry of GISAXS experiments.
A monochromatic X-ray beam with a wave vector $\vec{k}_i$ impinges on the sample surface at a grazing incidence angle $\alpha_i$.
The elastically scattered wave vector $\vec{k}_f$ propagates along the exit angle $\alpha_f$ and the azimuthal angle $\theta_f$.
The sample can be rotated around the $z$-axis by the angle $\varphi$.}
\label{fig:gisaxs_geometry}
\end{SCfigure*}

The measurement geometry of GISAXS \cite{levine_1989_GrazingincidenceSmallangle} is shown schematically in figure \ref{fig:gisaxs_geometry}.
The sample is illuminated under grazing incidence angle $\alpha_i$, and the resulting reflected and scattered radiation is collected with an area detector at exit angles $\alpha_f$ and $\theta_f$.
We chose our coordinate system such that the $x$-$y$-plane is the sample plane and the $x$-axis lies in the scattering plane, with the $z$-axis perpendicular to the sample plane.
In this coordinate system, the scattering vector $\vec{q} = \vec{k}_f - \vec{k}_i$ takes the form
\begin{align}
q_x &= k (\cos \theta_f \cos \alpha_f - \cos \alpha_i) \label{eq:q} \\
q_y &= k (\sin \theta_f \cos \alpha_f)   \\
q_z &= k (\sin \alpha_i + \sin \alpha_f) 
\end{align}
with the wave vector of the incoming beam $\vec{k}_i$, the wave vector of the scattered beam $\vec{k}_f$, $k = |\vec{k}_i| = |\vec{k}_f| = 2 \pi / \lambda$ and the wavelength $\lambda$ of the incident radiation.

\subsection{Experimental Setup}
The experiments were conducted at the four-crystal monochromator (FCM) beamline~\cite{krumrey_2001_HighaccuracyDetector} in the laboratory~\cite{beckhoff_2009_QuartercenturyMetrology} of the Physikalisch-Technische Bundesanstalt (PTB) at the electron storage ring BESSY~II. This beamline allows the adjustment of the photon energy in the range from \SI{1.75}{keV} to \SI{10}{keV}.
The beam spot size was about \SI{0.5x0.5}{mm} at the sample position, with an estimated vertical coherence length projected onto the sample in the $x$-direction of $\epsilon_x = \SI{100}{\micro\meter}$.
The GISAXS setup at the FCM beamline consists of a sample chamber \cite{fuchs_1995_HighPrecision} and the Helmholtz-Zentrum Berlin (HZB) SAXS setup \cite{gleber_2010_TraceableSize}.
The sample chamber is equipped with a goniometer which allows sample movements in all directions with a resolution of \SI{3}{\micro\meter} as well as rotations around all sample axes with an angular resolution of \ang{0.001}.
The HZB SAXS setup allows the movement of the in-vacuum Pilatus~1M area detector \cite{wernecke_2014_CharacterizationInvacuum}, reaching sample-to-detector distances from about \SI{2}{m} to about \SI{4.5}{m} and exit angles $\alpha_f$ up to approximately \ang{2}.
Along the whole beam path including the sample site, a high vacuum (pressure below \SI{e-8}{mbar} up to and including the sample site, pressure below \SI{e-4}{mbar} between sample and detector) is maintained.

The measurements of the homogeneity of the \emph{EBL} and \emph{NIL hexagonal} samples were conducted at a photon energy of \SI{6}{keV}.
All other measurements were conducted at a photon energy of \SI{3}{keV}.
The distance from the sample to the detector was calibrated by triangulation using the position of the direct and specularly reflected beam on the detector at various detector positions (see supplementary material for details).
The incident ($\alpha_i$) and exit ($\alpha_f$) angles were calibrated by first aligning the sample parallel to the incident X-ray beam ($\alpha_i = 0$) and then using the position of the specularly reflected beam on the detector at $\alpha_i \neq 0$, which together with the distance from the sample to the detector yields the exact values for $\alpha_f$ and $\theta_f$.
The orientation of the crystallographic direction to the x-ray beam $\varphi$ was calibrated by tuning $\varphi$ until the diffraction pattern was symmetric along the specular axis, yielding $|\varphi| < \SI{0.01}{\degree}$.
For the homogeneity measurements, the angles were calibrated using the position of the attenuated direct beam taken without sample in the beam path and the specularly reflected beam on the detector, giving higher precision in the comparison of the different samples.
The raw GISAXS data and used analysis scripts producing the graphics in this paper are available in the supplementary material.

\subsection{Theoretical Description}
\label{ssec:theo_descr}

\begin{figure*}[t]
\includegraphics[width=\textwidth]{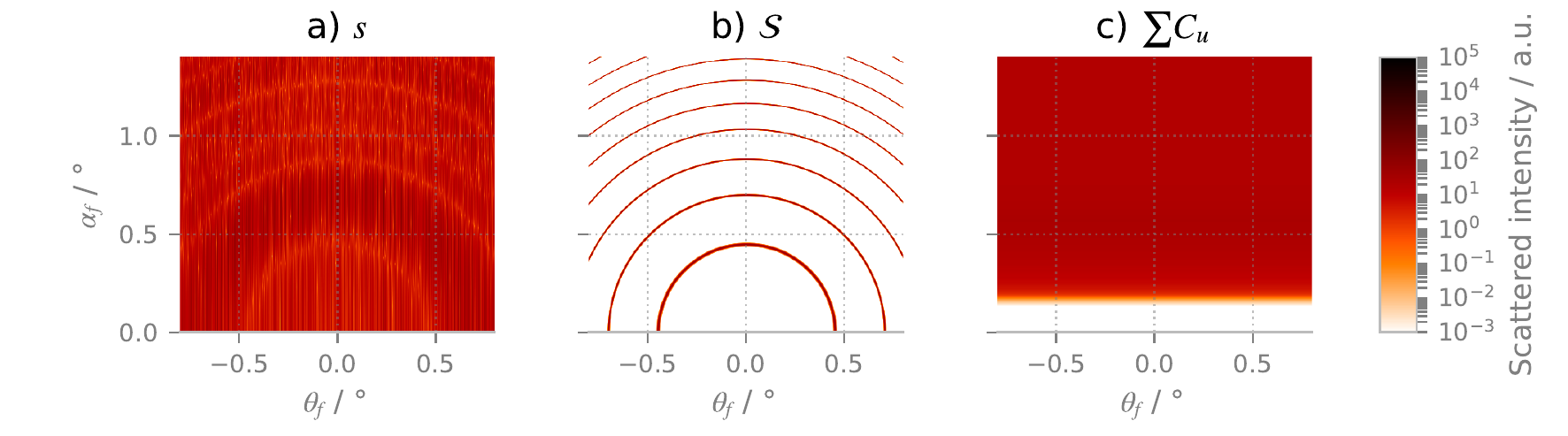}
\caption{Components of the distorted wave Born approximation model.
\textbf{a)} Interference $s$ within the locally quasicrystalline supercell.
\textbf{b)} Tiling interference function $\mathcal{S}$.
\textbf{c)} Trivial DWBA form factor $\mathcal{F}=\sum_{u=1}^4 C_u$ derived from the field amplitudes calculated from the layer system.}
\label{fig:dwba_components}
\end{figure*}

We model the GISAXS measurements using the framework of the distorted wave Born approximation (DWBA) \cite{vineyard_1982_GrazingincidenceDiffraction,sinha_1988_XrayNeutron,renaud_2009_ProbingSurface}. In the DWBA, the scattering cross section of a collection of particles is expressed as
\begin{align}
\frac{d\sigma}{d\Omega} &= \left|\sum_j \, e^{i \vec{q} \vec{r}_j} \FF_j(\vec q) \right|^2 \quad ,
\end{align}
with the position of the $j$-th particle $\vec r_j$ and its DWBA form factor $\FF_j$ \cite{burle_2016_BornAgain,daillant_2009_XrayNeutron}.

To describe the periodic samples, we use the \emph{decoupling approximation} for the pillars \cite{burle_2016_BornAgain}, assuming that all pillars have the same DWBA form factor $\FF_j = \FF$. This allows us to separate the description of the form factor and the interference function $\CS$
\begin{align}
\frac{d\sigma}{d\Omega} &= \CS(\vec q) \left|\FF(\vec q) \right|^2 \quad .
\end{align}
The interference function of a two-dimensional periodic lattice is
\begin{align}
\CS(\vec q) &= 4 \pi \sum_{\vec q_i \in \Lambda} \delta(\vec q - \vec q_i) \quad,  \label{eq:interference_periodic_twodlattice}
\end{align}
with the reciprocal lattice $\Lambda$ \cite{burle_2016_BornAgain}.

To describe the locally quasicrystalline samples, we separately describe the positions of the $N$ nanopillars in the supercell by explicitly enumerating each position within the supercell, and fill the plane by tiling supercells.
In effect, we use the model of a two-dimensional crystal with a large unit cell.
Figure~\ref{fig:dwba_components} shows the different effects of the pillar positions within the supercell, the supercell tiling, and the form factor of the individual pillars.
The calculations for figure~\ref{fig:dwba_components} were done with a photon energy of \SI{3}{keV}, the corresponding refractive index of silicon oxide according to \citeasnoun{henke_1993_XrayInteractions} and an incident angle of $\alpha_i=\SI{0.7}{\degree}$.

The tiling of the supercells is described by an interference function $\CS$ again using the \emph{decoupling approximation} between supercells \cite{burle_2016_BornAgain}, stating that pillar properties in the same sample do not depend on the specific supercell the pillars are in.
This leads to the expression
\begin{align}
\frac{d\sigma}{d\Omega} &= \left| \CS(\vec q) \sum_j^N \, e^{i \vec q \vec r_j} \FF_j(\vec q) \right|^2 \, .
\end{align}
If we further assume that all pillars in a sample in each supercell have the same form and thus a uniform DWBA form factor $\FF_j = \FF$, we can further simplify the expression as follows
\begin{align}
\frac{d\sigma}{d\Omega} &= \left| \CS(\vec q) \FF(\vec q) \sum_j^N \, e^{i \vec q \vec r_j} \right|^2 \\
                        &= \left| \CS(\vec q) \FF(\vec q) s(\vec q) \right|^2 \, , \label{eq:scattering_cross_section}
\end{align}
where $s(\vec q)$ contains only the effect of the positions of pillars in the locally quasicrystalline supercell and is given by the positions of the nanopillars known from the chosen design.
An example of a result for $s(\vec q)$ is shown in figure~\ref{fig:dwba_components}a).
A complex pattern of diffraction orders which are sharp in $\theta_f$ and elongated in $\alpha_f$ can be seen.
The distribution of the diffraction orders in the $\theta_f$-direction arises from the local correlations in the $y$-direction.
In contrast, the distances between individual pillars cannot be resolved in the $x$-direction, but the size of the rectangular supercell can be resolved and manifests itself in diffuse semicircles at constant $q_x$ that overlay the scattering pattern.

\begin{figure*}[t]
\includegraphics[width=\textwidth]{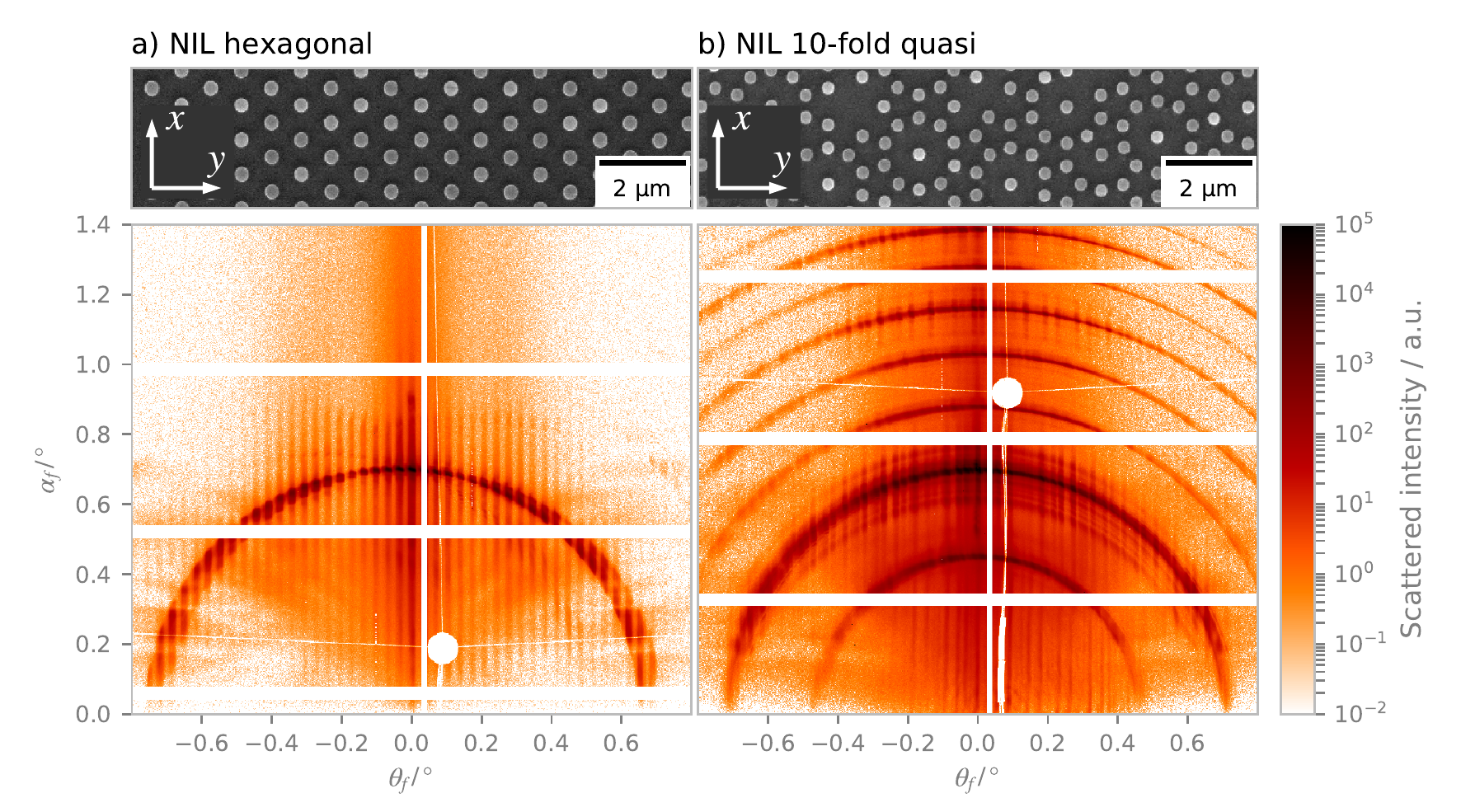}
\caption{Scanning electron microscopy (top) and GISAXS (bottom) measurements of \textbf{a)} the \emph{NIL hexagonal} and \textbf{b)} \emph{NIL 10-fold quasi} samples. Solid white areas in the GISAXS measurements are due to detector gaps and a beam stop. GISAXS measurements were taken with the incident beam aligned to the crystallographic direction, resulting in the main intensity being scattered into a semicircle of diffraction orders. Additional semicircles in the measurement of the \emph{NIL 10-fold quasi} sample show the periodicity of the supercell in its pattern, with a period length of about \SI{9.5}{\micro\meter}.}
\label{fig:gisaxs_c_vs_qc}
\end{figure*}

To be able to implement eq.~\eqref{eq:scattering_cross_section} numerically, we have to find explicit expressions for $\CS$ and $\FF$ as well.
For the tiling interference function $\CS$, we use a one-dimensional lattice interference function in the $x$-direction with a Gaussian decay function with decay length $\lambda_d$ \cite{burle_2016_BornAgain}, leading to 
\begin{align}
\CS(q_x) &= \frac{\sqrt{2\pi} \lambda_d}{p} \sum_{n \in \ZZ} e^{-(q_x - 2 \pi n / p)^2 \lambda_d^2 / 2} \quad ,
\end{align}
with the tiling period $p$ (see fig.~\ref{fig:dwba_components}b).
In practice, due to the large angular distance between peaks in $q_x$, it is sufficient to calculate the largest contribution with a single $n$ to the infinite sum, with $n = \round(q_x p / (2 \pi))$.
The tiling in the $y$-direction can be neglected since the experimental resolution is not high enough to resolve the very dense spacing of peaks in $q_y$.
For the DWBA form factor $\FF$, we use the general form \cite{burle_2016_BornAgain,lazzari_2009_GrazingIncidence}
\begin{align}
\FF(\vec q) &= \sum_{u=1}^4 \, C_u F(\vec q_u) \, \\
\textrm{with} & \nonumber \\
C_1 &= A_f^- \, A_i^- & \vec q_1 &= \vec k_f^- - \vec k_i^- \\
C_2 &= A_f^- \, A_i^+ & \vec q_2 &= \vec k_f^- - \vec k_i^+ \\
C_3 &= A_f^+ \, A_i^- & \vec q_3 &= \vec k_f^+ - \vec k_i^- \\
C_4 &= A_f^+ \, A_i^+ & \vec q_4 &= \vec k_f^+ - \vec k_i^+ \, ,
\end{align}
where $A_{f,i}^{\pm}$ are the electric field amplitudes of the upwards (+) and downwards (-) travelling waves of the incoming (i) and final (f) waves as obtained from the dynamical calculation of the layer system without particles \cite{gibaud_2009_SpecularReflectivity} and $k_{f,i}$ are the respective wave vectors.
Since we are mainly interested in the locally quasicrystalline nature of the samples and not in the form of the pillars, we can  set the Born form factor $F(\vec q)$ to a constant (see fig.~\ref{fig:dwba_components}c).
Finally we get:
\begin{align}
\frac{d\sigma}{d\Omega} (\vec q) \propto & \left| \CS(q_x) \, s(\vec q) \, \FF(\vec q) \right|^2 \\
\propto & \left| \frac{\sqrt{2\pi} \lambda_d}{p} e^{-(q_x - 2 \pi n / p)^2 \lambda_d^2 / 2} \right. \nonumber \\
&\left. \sum_j^N \, e^{i \vec q \vec r_j} \sum_{u=1}^4 C_u(\vec q_u) F(\vec q_u) \right|^2 \, . \label{eq:final_scattering_cross_section}
\end{align}

We have implemented eq. \eqref{eq:final_scattering_cross_section} in \emph{Python} \cite{millman_2011_PythonScientists} with the \emph{numpy} and \emph{numba} \cite{lam_2015_NumbaLLVMbased} packages.
The implementation is available in the supplementary material to this paper.

\section{Results}
\subsection{Comparison of periodic and locally quasiperiodic structures}

GISAXS measurements of the \emph{NIL hexagonal} and \emph{NIL 10-fold quasi} samples collected at an incident angle of $\alpha_i = \ang{0.7}$ and a sample-to-detector distance of \SI{4.527}{\meter} are shown in figure \ref{fig:gisaxs_c_vs_qc}.
We aligned the crystallographic direction of the samples to the incident X-ray beam, as can be seen from the mirror symmetry of the diffraction patterns.
The resolution of the measurements is limited by the divergence of the incident beam, and is $\Delta q_y \approx \SI{0.005}{\per\nano\meter}$.

\begin{SCfigure*}[1.0][t]
\includegraphics[width=\columnwidth]{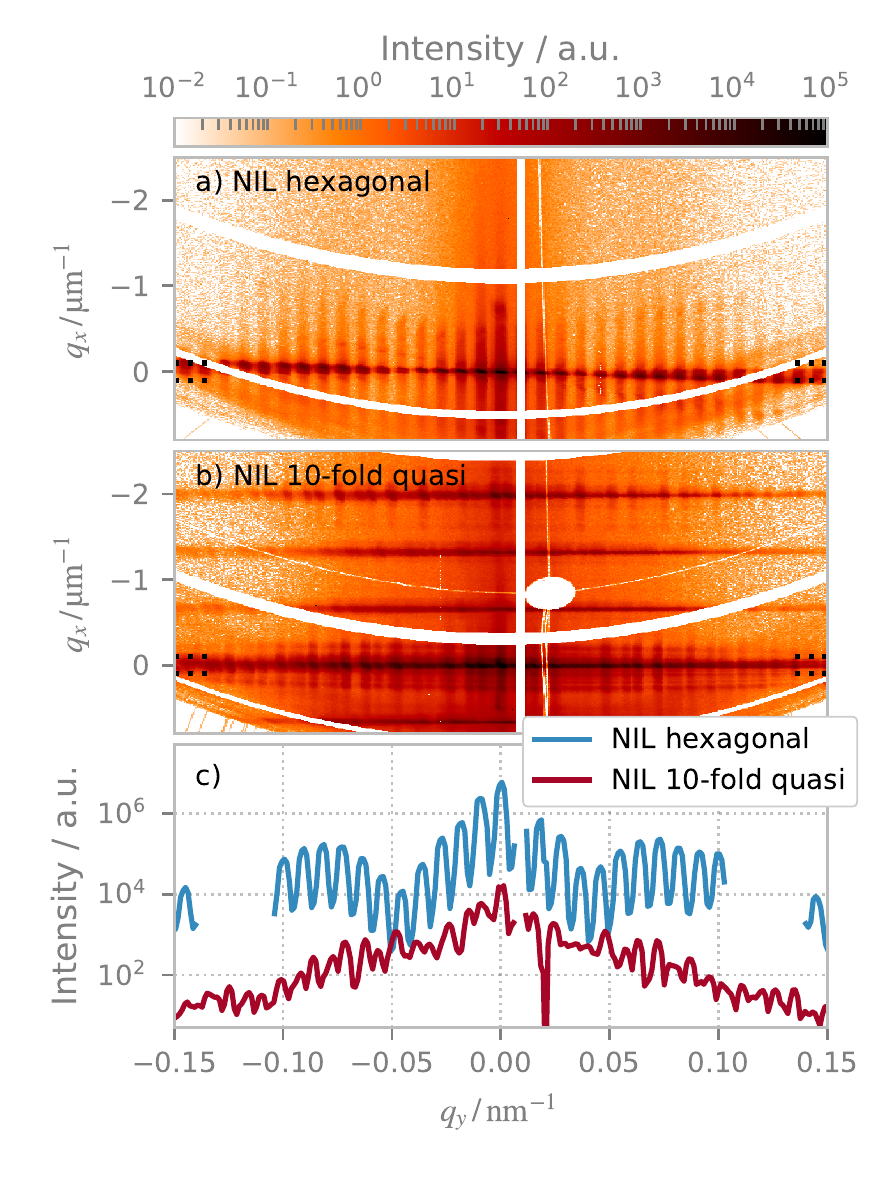}
\caption{GISAXS measurements of the \textbf{a)} \emph{NIL hexagonal} and \textbf{b)} \emph{NIL 10-fold quasi} sample projected onto the $q_y$-$q_x$-plane. Note the different resolution in $q_y$ and $q_x$.
In \textbf{a)}, scattering is confined to $q_x \approx 0$, while \textbf{b)} shows scattering at $q_x \approx n \, \SI{0.66}{\per\micro\meter}$ with integer $n$ due to the supercell periodicity.
\textbf{c)} shows cuts at $\SI{-0.1}{\per\micro\meter} < q_x < \SI{0.1}{\per\micro\meter}$ as indicated by the dashed lines at the sides of the scattering patterns.
The periodicity in the scattering of the \emph{NIL hexagonal} sample is readily visible, while the \emph{NIL 10-fold quasi} sample shows no periodicity in $q_y$.
The cut of the \emph{NIL 10-fold quasi} sample was shifted for visibility.}
\label{fig:gisaxs_c_vs_qc_qx}
\end{SCfigure*}

The scattering pattern of the \emph{NIL hexagonal} sample (fig. \ref{fig:gisaxs_c_vs_qc}a) consists of diffraction orders forming a semicircle.
The pattern can be described by the intersection of the reciprocal form of a hexagonal lattice with the Ewald sphere of elastic scattering (similarly to the reciprocal space construction of the scattering of line gratings as done by \citeasnoun{mikulik_2001_CoplanarNoncoplanar} and \citeasnoun{yan_2007_IntersectionGrating}).
The reciprocal form of a hexagonal lattice with lattice constant $a$ is a collection of lattice truncation rods that are parallel to the $q_z$-axis and are arranged on a hexagonal lattice with lattice constant $\frac{4\pi}{a\sqrt{3}}$ in the $q_x$-$q_y$-plane.
When the incident beam is aligned to a symmetry axis of the hexagonal lattice, the reciprocal form of the hexagonal lattice is aligned to the $q_x$-axis, with lattice truncation rods at $q_x=\frac{n_x 2\pi}{a}$ with the order $n_x$.
Intersecting the Ewald sphere with the lattice truncation rods yields diffraction peaks.
The steep inclination of the Ewald sphere at small exit angles results in a small covered $q_x$-range in GISAXS measurements.
Therefore, only the diffraction peaks resulting from the zero-order lattice truncation rods at $q_x=0$ can be measured, leading to a semicircle of diffraction peaks, as seen in the scattering pattern.
However, in contrast to the theory of a perfect hexagonal lattice, the diffraction orders are stretched in $\alpha_f$ at higher $\theta_f$, which we attribute to inhomogeneities within the sample pattern \cite{rueda_2012_GrazingincidenceSmallangle,meier_2012_SituFilm,soccio_2014_AssessmentGrazingincidence} and discuss in detail in subsection \ref{ssec:homogeneity}.

The diffraction pattern of the \emph{NIL 10-fold quasi} sample (fig. \ref{fig:gisaxs_c_vs_qc}b) shows similarities and striking differences to the diffraction pattern of the \emph{NIL hexagonal} sample.
As in the diffraction of the periodic sample, there are diffraction orders forming a semicircle and the diffraction orders are stretched along $\alpha_f$ due to inhomogeneities (in the example shown in figure \ref{fig:gisaxs_c_vs_qc}b, the inhomogeneities are relatively small; a detailed description of the inhomogeneities is found in section \ref{ssec:homogeneity}).
However, there are also two main differences.
Firstly, additional, weaker semicircles appear above and below the semicircle at $q_x=0$.
They are caused by the long-range periodicity of the supercell design (compare section \ref{ssec:theo_descr}), with a period length of about \SI{9.5}{\micro\meter} in the $x$-direction.
Secondly, the diffraction orders show no periodicity within the semicircle, which is due to the short-range quasiperiodicity in the $y$-direction.
The long-range periodicity in the $y$-direction cannot be resolved.

The periodicity in the $x$-direction is more clearly visible after transforming the measurements according to eq. \eqref{eq:q} into the $q_y$-$q_x$-plane, shown in figure \ref{fig:gisaxs_c_vs_qc_qx}a) and figure \ref{fig:gisaxs_c_vs_qc_qx}b).
The semicircles are transformed into straight lines spaced equally in $q_x$.
The intensity distribution along the $q_y$-direction is readily visible in in-plane cuts taken by integrating along $q_x$ at $\SI{-0.1}{\per\micro\meter} < q_x < \SI{0.1}{\per\micro\meter}$, shown in figure \ref{fig:gisaxs_c_vs_qc_qx}c).
While the measurement of the periodic sample shows equally spaced diffraction orders, the intensity of the locally quasiperiodic sample does not display a discernible periodicity.

\begin{figure}[tb]
\includegraphics[width=\columnwidth]{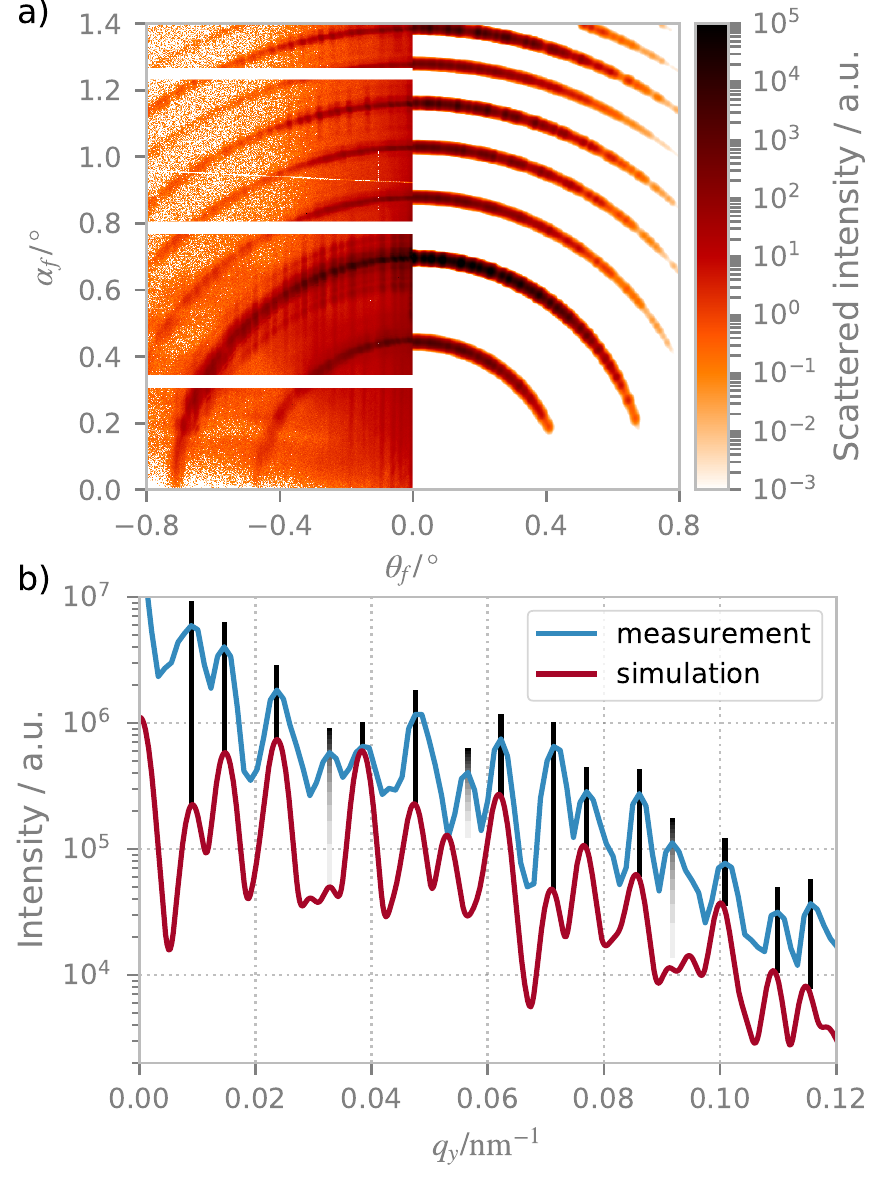}
\caption{\textbf{a)} Comparison of GISAXS measurements (left) and DWBA simulation (right) of the \emph{NIL 10-fold quasi} sample.
\textbf{b)} Comparison of $q_y$-cuts taken along the semicircle around $q_x=0$.
Black lines indicate measured peak positions for easier comparison with the simulated data.
Simulated data has been shifted downwards for clarity.}
\label{fig:gisaxs_qc_simulation_vs_measurement}
\end{figure}

We simulated the diffraction of the \emph{NIL 10-fold quasi} sample as detailed in subsection \ref{ssec:theo_descr}.
For the tiling period, we use the design value of $p=\SI{9.4}{\micro\meter}$ and set the decay length to $\lambda_d = \SI{200}{\micro\meter}$.
For the form factor $F$, we use a constant and to calculate the DWBA prefactors $C_u$, we assume a system consisting of three layers.
At the top is a vacuum layer, at the bottom, the silicon oxide substrate and between them, an average density layer generated by the nanopillars.
The average density layer also consists of silicon oxide with parameters according to the nanopillar design values.
Its density is reduced to $0.1$ relative to bulk silicon oxide and the thickness of the layer is \SI{275}{\nano\meter}, the nominal height of the pillars.
For the comparison of the simulated data with the measurement, a Debye-Waller factor $e^{-q_y^2 \, \sigma_{rms}^2}$ with a mean roughness $\sigma_{rms} = \SI{18}{nm}$ was introduced in the simulation. Additionally, to account for the beam divergence in the experiment, a Gaussian filter with a standard deviation of \SI{0.005}{\degree} was applied to the simulation.
A comparison of the measurement and simulation results for the \emph{NIL 10-fold quasi} sample is shown in figure \ref{fig:gisaxs_qc_simulation_vs_measurement}.
The position and form of the semicircles match very well (fig. \ref{fig:gisaxs_qc_simulation_vs_measurement}a). In the simulation, the position and form of the semicircles results from the tiling of the supercells which is described in section \ref{ssec:theo_descr} by $\CS(\vec q)$.
To also assess the accuracy of our description of the locally quasicrystalline ordering within the supercells (described by $s(\vec q)$), we took a cut at $q_x=0$, which is shown in figure \ref{fig:gisaxs_qc_simulation_vs_measurement}b).
The positions of the peaks agree well with only three peaks missing or shifted in the simulation, but the relative peak intensities match poorly.
The comparison between the simulation and the measurement shows that our hierarchical model of tiled supercells with local quasicrystalline ordering can adequately describe the position of the coherent scattering features, but is not able to describe their relative intensity satisfactorily.

\begin{figure}[tb]
\includegraphics[width=\columnwidth]{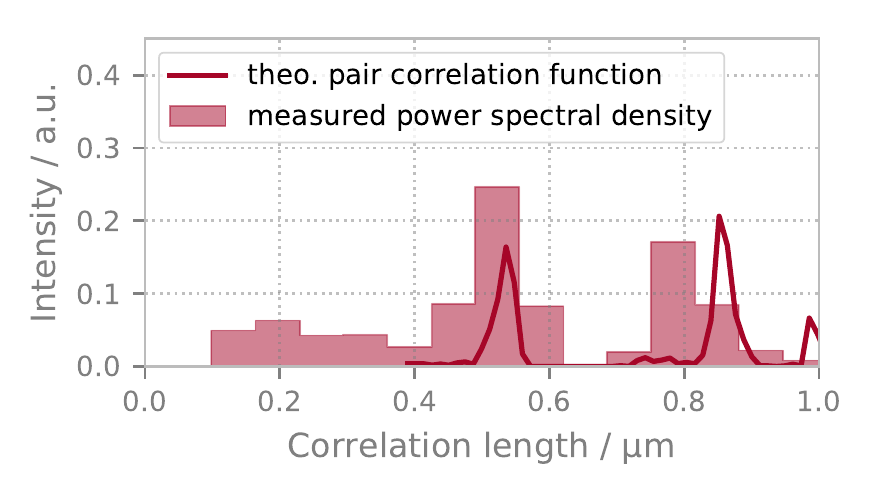}
\caption{Power spectral density of the GISAXS $q_y$-cuts at $q_x = 0$ and pair correlation function of the design lattice of the \emph{NIL 10-fold quasi} sample.}
\label{fig:fourier_c_vs_qc}
\end{figure}

To obtain further insights in real space, we calculated the power spectral density of the in-plane cuts using the range of $\SI{-0.1}{\per\nano\meter} < q_y < 0$, which avoids all beam stop shadowing and detector gaps.
The discrete Fourier transformation (DFT) was carried out using a Kaiser window \cite{kaiser_1966_DigitalFilters}.
For comparison with theoretical expectations, we computed the pair correlation function of the design pillar positions.
The result for the \emph{NIL 10-fold quasi} sample is shown in figure \ref{fig:fourier_c_vs_qc}.
The measured power spectral density shows two broad peaks at about \SI{0.55}{\micro\meter} and \SI{0.8}{\micro\meter}.
Correspondingly, the theoretical pair correlation function of the design lattice shows two main peaks, at about \SI{0.55}{\micro\meter} and \SI{0.85}{\micro\meter}.
The main peaks at about \SI{0.5}{\micro\meter} and \SI{0.8}{\micro\meter} agree quite well with the measurement and theoretical expectation and correspond to the nearest neighbour distance and the ring-like patterns in the sample, respectively (compare SEM image in figure \ref{fig:gisaxs_c_vs_qc}b).
However, a non-zero intensity between approximately \SI{0.1}{\micro\meter} and \SI{0.4}{\micro\meter} is not expected theoretically.
We attribute this to the finite width of the pillars, which was not considered in the pair correlation function.

\subsection{Lattice distortions}
\label{ssec:homogeneity}

As noted, the diffraction patterns of the \emph{NIL} samples show an elongation of the diffraction orders along $q_z$ at high $q_y$.
In this section, we use this elongation to quantify the inhomogeneity of all four samples, scanning along the $y$-direction for the spatial resolution and characterization of the whole sample surface.

\begin{figure}[tb]
\includegraphics[width=\columnwidth]{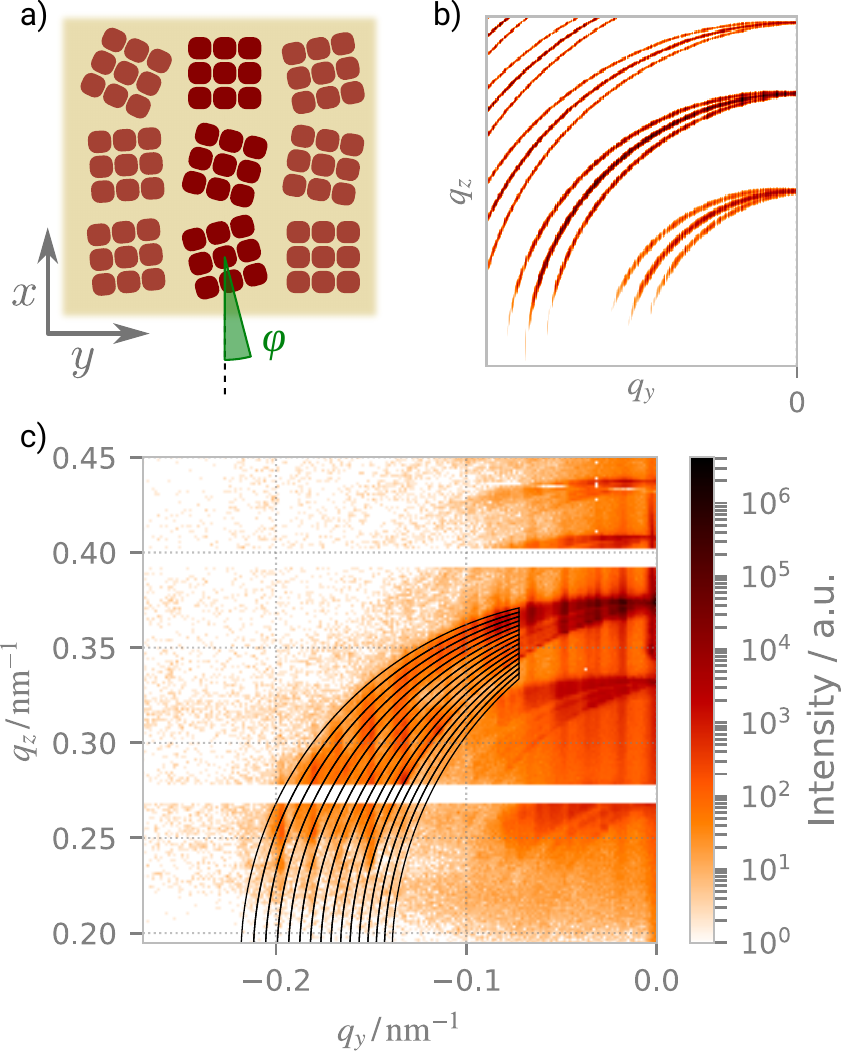}
\caption{\textbf{a)} Top view sketch of our sample model. We model the sample using \si{\milli\meter}-sized domains of perfectly ordered lattices of supercells, each having a slightly different azimuthal orientation $\varphi$.
\textbf{b)} Simulation of the model. Each domain leads to scattering into tilted semicircles, and the signals combine incoherently.
The diffraction along the semicircles overlaps, leading to elongation in the $q_z$-direction and separation of the semicircles at higher $q_y$.
\textbf{c)} Measurement of $\varphi$ distribution. For the measurement of $\varphi$ variations, the GISAXS image is cut into slices corresponding to a $\varphi$ range, and the intensity in each slice is integrated.
In the analysis, the slices are twice as dense as shown here.
This example shows scattering of the \emph{NIL 12-fold quasi} sample.}
\label{fig:phi_var_sketch_combined}
\end{figure}

In section \ref{ssec:theo_descr}, we use the tiling interference function $\CS(q_x)$ of a one-dimensional lattice in the $x$-direction, which leads to scattering into a series of sharp arcs.
To model the disturbed order in the locally quasiperiodic samples, we assume that the tiling is locally well ordered, but due to the manufacturing, several domains exist which have the tiling in a slightly different direction and therefore, different $\varphi$ (see fig. \ref{fig:phi_var_sketch_combined}a).
Equivalently, we assume, for the periodic samples, that the structure is locally well ordered, but several domains exist which are oriented slightly differently.
The domains are assumed to be large compared to the coherence length of the incident X-rays, and therefore the measured signal is the incoherent superposition of the signals of the domains illuminated by the X-ray beam (see fig. \ref{fig:phi_var_sketch_combined}b).
To describe the signal of a single domain rotated in the sample plane by the angle $\varphi$, we generalize $\CS$. For the zero-order semicircle, this yields \cite{mikulik_2001_CoplanarNoncoplanar,yan_2007_IntersectionGrating,pfluger_2017_GrazingincidenceSmallangle} in coordinate form:
\begin{align}
q_y =& 2 \pi \cos(\varphi) b \label{eq:slicing_qy} \\
q_z =& 2 \pi / \lambda \biggl(\sin(\alpha_i) + \nonumber \\
     & \left. \sqrt{\sin^2(\alpha_i) - b^2 \lambda^2 - 2 \sin(\varphi) \cos(\alpha_i) b \lambda} \right) \label{eq:slicing_qz} \quad ,
\end{align}
with the running auxiliary variable $b$.

\begin{SCfigure*}[1.0][tb]
\includegraphics[width=1.3\columnwidth]{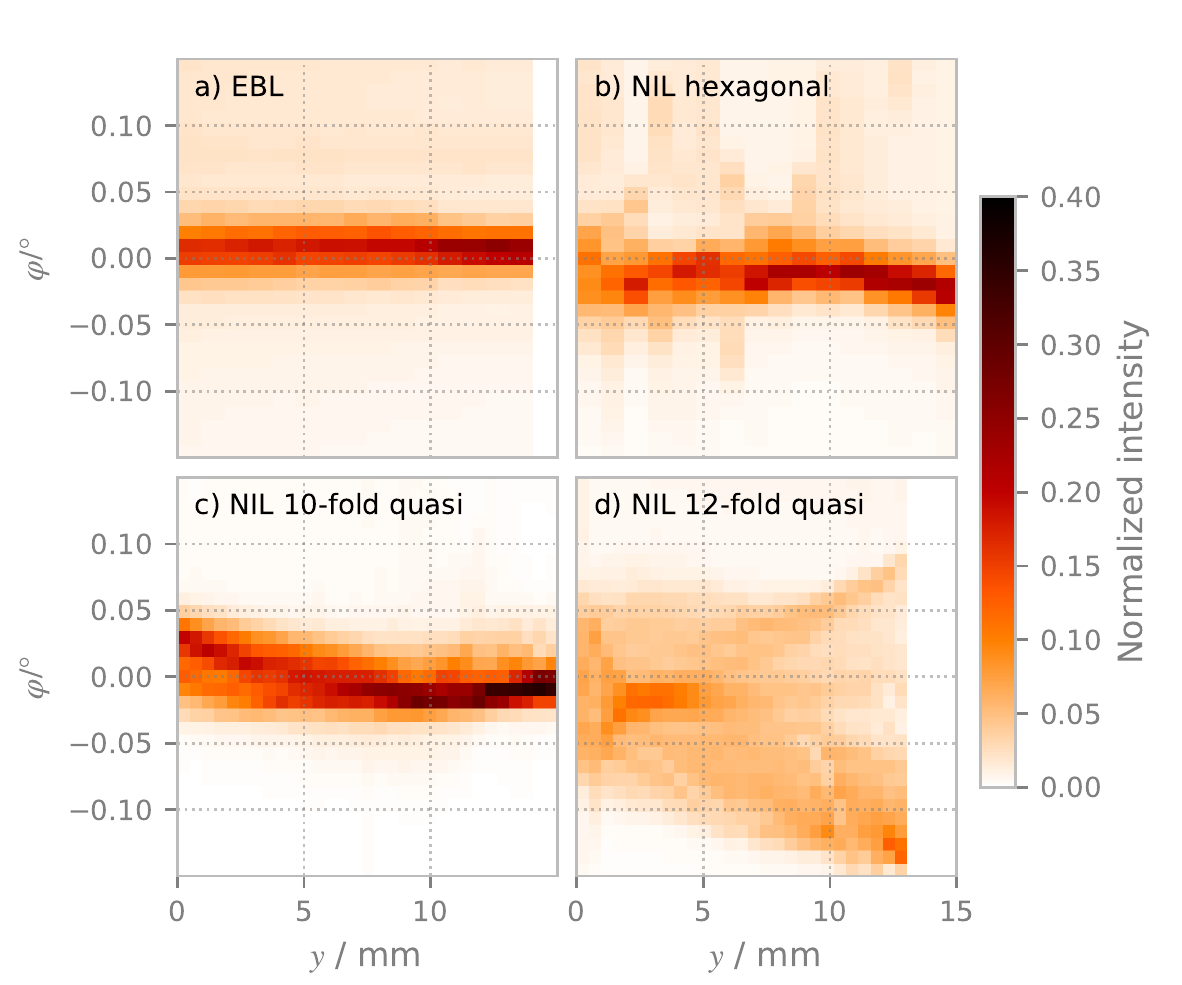}
\caption{$\varphi$ variation scanned along the $y$-direction.
The \emph{EBL} sample (a) shows a narrow, homogeneous $\varphi$ distribution with a width close to the resolution limit.
The \emph{NIL} samples all show larger inhomogeneities.
While the \emph{NIL hexagonal} and \emph{NIL 10-fold quasi} samples (b, c) have a rather narrow distribution, they display a drift of the central $\varphi$ along $y$.
Finally, the \emph{NIL 12-fold quasi} sample (d) shows a comparatively wide and inhomogeneous $\varphi$ distribution.
An interactive animation presenting diffraction patterns for the selected $y$ position on the \emph{NIL 12-fold quasi} sample is available in the supplementary material.}
\label{fig:phi_detail}
\end{SCfigure*}

Using eqs. \eqref{eq:slicing_qy} and \eqref{eq:slicing_qz}, the scattering image is cut into slices with differing $\varphi$ and the intensity in each slice is integrated (see fig. \ref{fig:phi_var_sketch_combined}c).
For the slicing of the scattering image, a trade-off between signal intensity and resolution needs to be made.
For high signal intensities, the slices need to be continued towards $q_y = 0$, because scattering at smaller $q$ is generally more intense due to roughness.
On the other hand, the height of the slices decreases towards $q_y = 0$.
Given a constant resolution in $q_z$ due to divergence or pixel size, extending the slices towards smaller $q_y$ will decrease the $\varphi$ resolution.
With the resolution in the exit angle $\alpha_f$ of our setup of \ang{0.005} and our slicing range of $q_y<-\SI{0.072}{\per\nano\metre}$ , we have a $\varphi$ resolution of about \ang{0.015} with acceptable signal-to-noise ratio.

We took measurements of the three samples produced by nanoimprint lithography (\emph{NIL hexagonal}, \emph{NIL 10-fold quasi} and \emph{NIL 12-fold quasi}) and of the \emph{EBL} sample as a comparison.
Since the elongated beam footprint is longer than the sample, each GISAXS measurement collects information on an approximately \SI{0.5}{mm} wide, full-length strip of the \SI{20x20}{mm} large sample.
To gain insight into the spatial distribution of $\varphi$ inhomogeneities, GISAXS measurements were taken in a scan along the $y$-direction.
The results are given in figure \ref{fig:phi_detail} and show the statistical inhomogeneities of each sample in an area of about \SI{15x20}{mm}.

The \emph{EBL} reference sample shows a very narrow $\varphi$ distribution, with no changes over $y$.
Apart from demonstrating the high quality of the E-beam lithography production process, the absence of any measurable drift also shows the angular stability of the measurement setup for sample movements.
In contrast, the nanoimprinted samples all display larger inhomogeneities.
Two (\emph{NIL hexagonal} and \emph{NIL 10-fold quasi}) show only slightly wider $\varphi$ distributions, which however drift along $y$.
The \emph{NIL 12-fold quasi} sample shows the largest inhomogeneities, both as wider $\varphi$ distributions at each $y$ position and as a high shift along $y$.

\begin{figure}[tb]
\includegraphics[width=\columnwidth]{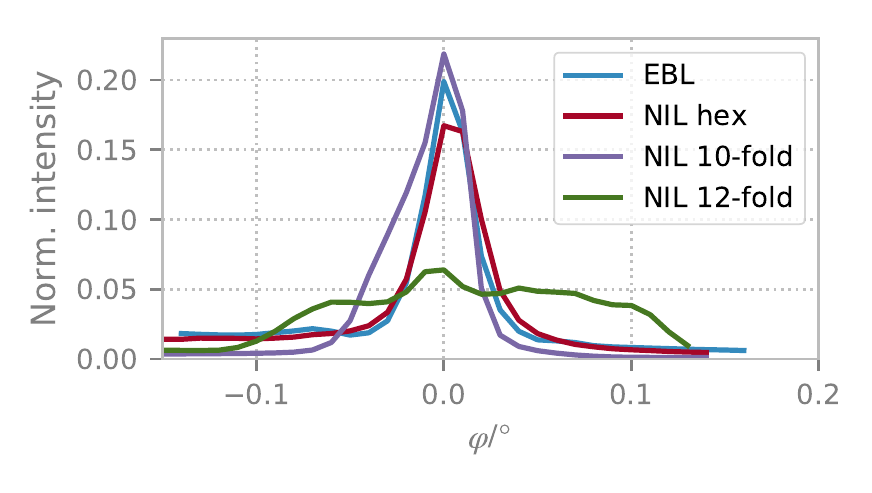}
\caption{Summary of $\varphi$ variations.
For each sample, the GISAXS measurements at all $y$ positions were summed up and the resulting distribution was centred at the maximum to give the relative $\varphi$ deviation.}
\label{fig:phi_summary}
\end{figure}

The results are condensed in figure \ref{fig:phi_summary}, which presents the sum of all measurements along $y$ for each sample.
It shows the very high quality of the E-beam lithography process, and the loss of homogeneity in the additional nanoimprinting processing steps.
All three \emph{NIL} samples were simultaneously manufactured on one substrate using one template, but they show differing $\varphi$ homogeneity.
The differences in $\varphi$ homogeneity cannot be explained by the different lattice design, since each \emph{NIL} sample shows large spatial variations in $\varphi$ homogeneity in the scan in the $y$-direction.
We conclude that inhomogeneities were introduced in the nanoimprinting process, which can be explained with the flexible NIL template, which leads to small variations in the imprint.
The absolute magnitude of the largest $\varphi$ variations is still relatively small with $\varphi = \pm \ang{0.15}$.
As explained in section \ref{ssec:samples}, the nanoimprinting process was not optimized for the highest homogeneity.

\section{Conclusion}
To our knowledge, we report on the first GISAXS measurement of locally quasiperiodic surface structures, and compare them with simulations and with measurements of periodic samples.
The measurements agree qualitatively with the theoretical description of the locally quasiperiodic samples as a hierarchically ordered system.
The supercell gives rise to quasicrystalline diffraction at small length scales, and the tiling of the rectangular supercells leads to crystalline diffraction at longer length scales.
Cuts of the GISAXS measurements in the $q_y$-direction and their Fourier transform show the local quasiperiodicity and agree with the simulation and the calculated pair correlation function, respectively.
Simultaneously, the long-range periodicity is revealed by GISAXS in the $q_x$-direction due to the \si{\micro\meter}-resolution in $x$.
In contrast, the measurements of periodic samples show simple diffraction patterns arising from the short-range periodicity visible in $q_y$.

With our theoretical description of rotated lattices, we are able to extract the distortions of the periodic supercell tiling of the locally quasicrystalline samples and the distortions of the periodic lattice of the periodic samples.
We extract the lattice distortions on a \SI{15x20}{mm} area spatially resolved in the $y$-direction.
Because GISAXS is very sensitive to misalignments of the local lattice direction, we can resolve distortions down to \ang{0.015}.
As expected, the \emph{EBL} sample shows minimal lattice distortions over the whole sample, demonstrating the high quality achievable using electron beam lithography.
Likewise, we find only small lattice misalignments in the nanoimprinted samples considering their intended use in the solar spectrum.
However, in comparison to the \emph{EBL} sample, the variations in homogeneity are considerably larger, with lattice misalignments up to $\pm \ang{0.15}$.
By optimizing the nanoimprinting process, the homogeneity of the nanoimprinted samples could likely be increased further.

We have shown that GISAXS is a suitable method to investigate complex surface designs intended for use in solar cells, and an extension to buried interfaces is desirable.
However, to extract quantitative structure parameters of the investigated locally quasiperiodic samples from the GISAXS measurements, a more quantitative theoretical description would be needed.
In particular, we neglected the Born form factor of the individual pillars, focussing on the positions of the pillars, but the average form of the pillars is also of interest for the intended application.
However, for the reconstruction of the average form of the pillars, a more extensive dataset is likely necessary.

A comparison between locally quasiperiodic samples and globally quasiperiodic samples without long-range periodicity would be of special interest.
For such a comparison, an interference function describing an infinite quasiperiodic lattice would need to be developed to describe the globally quasiperiodic samples in the distorted wave Born approximation.
Unfortunately, due to manufacturing constraints, no such samples exist to our knowledge.

\section{Acknowledgments}
The authors wish to thank Levent Cibik and Stefanie Langner for their support during the experiments.

The German Ministry of Education and Research is acknowledged for funding the research activities of the Young Investigator Group Nano-SIPPE at HZB in the NanoMatFutur programme (no. 03X5520).

\bibliography{zotero}

\end{document}


\date{}
\title{Supporting Information \\ GISAXS calibration procedure}
\subtitle{Distortion analysis of crystalline and locally quasicrystalline 2D photonic structures with GISAXS}
\author{Mika Pflüger \and Victor Soltwisch \and Jolly Xavier \and Jürgen Probst \and Frank Scholze \and Christiane Becker \and Michael Krumrey}
\maketitle


\begin{figure}[hb]
\centering
\includegraphics{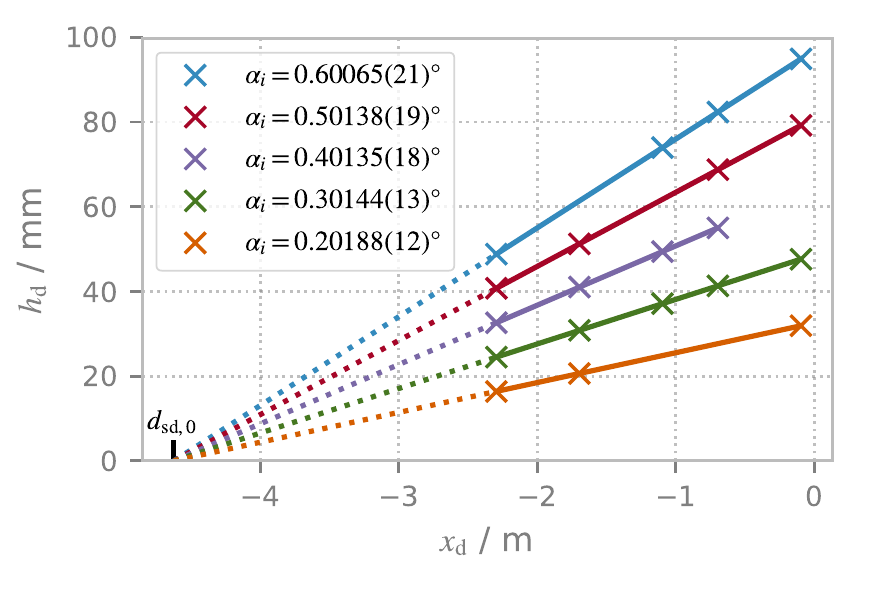}
\caption{Determination of the sample-to-detector distance offset $d_\mathrm{sd,0}$ and the incidence angle $\alpha_i$ by triangulation.
Crosses denote the measured data, the solid lines are the fit.
Dashed lines show the extension of the fit to the origin, which yields $d_\mathrm{sd,0}$.}
\label{fig:triangulation}
\end{figure}

For the GISAXS measurements, the distance between the sample and the detector $d_\mathrm{sd}$ is not known \emph{a priori}.
The position of the detector on the sledge $x_\mathrm{d}$ is measured using optical encoders, so that $d_\mathrm{sd} = x_\mathrm{d} - d_\mathrm{sd,0}$ with the sample detector distance offset $d_\mathrm{sd,0}$.
The measurement of $d_\mathrm{sd,0}$ is performed using triangulation with the specularly reflected beam.
For each detector position $x_\mathrm{d}$, the distance on the detector between the direct beam measured without a sample and the specularly reflected beam $h_\mathrm{d}$ is measured for $N$ incidence angles $\alpha_{i,n}$. The results are shown in figure \ref{fig:triangulation}.
The data are fitted to the model
\begin{align}
h_\mathrm{d} &= \tan(2\alpha_{i,n}) \, (x_\mathrm{d} - d_\mathrm{sd,0})
\end{align}
using the Levenberg-Marquardt method with the free parameters $\alpha_{i,n}$ and $d_\mathrm{sd,0}$.
The resulting fit is shown in figure \ref{fig:triangulation} as well.
Uncertainties of the fit parameters are estimated from the goodness of the fit using the covariance matrix of the Levenberg-Marquardt fit scaled by the reduced chi-square, for a final uncertainty of \SI{1.3}{mm} for the sample detector distance offset.